\documentstyle[aps,multicol,epsfig,prl]{revtex}

\input epsf

\def\prb{Phys.\ Rev.\ B}

\def\prl{Phys.\ Rev.\ Lett.\/}

\def\be{\begin{equation}}
\def\ee{\end{equation}}
\def\ba{\begin{eqnarray}}
\def\ea{\end{eqnarray}}

\def\248{Y$_2$Ba$_4$Cu$_8$O$_{7-\delta}$}

\def\C60{A$_x$C$_{60}$}

\begin{document}

\title
{Electron Fractionalization}

\author{S.~A.~Kivelson}
\address
{ Dept.\ of Physics, U.C.L.A., Los Angeles, CA  90095}

\date{\today}
\maketitle
\begin{abstract}
A largely qualitative, and rather idiosyncratic discussion of electron
fractionalization in condensed matter physics is presented, including some
historical reflections and some speculations concerning future application of
these ideas. Particular attention is paid to systems which exhibit
spin-charge separation, {\it i.e.} the electron can decay into separate
excitations which carry the electron spin and the electron charge;  the
soliton model of polyacetylene is treated as a paradigmatic example.  This
paper is based on a talk given at a symposium honoring A.J.Heeger,
A.G.MacDiarmid, and H.Shirakawa, the winners of the year 2000 Nobel Prize in
Chemistry.

\end{abstract}

\begin{multicols}{2}
\narrowtext

\section{Spin-Charge Separation in the 1DEG}

As far as I know, the idea of electron fractionalization rose
out of studies of the one-dimensional electron gas\cite{review} (1DEG).  In
particular, the idea of spin-charge separation was first explicitly treated by
Luther and Emery\cite{le} in the context of a continuum (field theory) limit
of the 1DEG.  What they showed is that the Hamiltonian can
be expressed as a sum
\be
H_{1DEG}=H_c[\phi_c]+H_s[\phi_s]+H_{irr}[\phi_c,\phi_s]
\ee
where $H_c$ and $H_s$ are, respectively, the Hamiltonians which govern the
dynamics of the spin and charge fields, $\phi_c$ and $\phi_s$, respectively,
and $H_{irr}$ (which they did not treat, explicitly) consists of terms that can
be neglected in the long wave-length limit (irrelevant terms, in the
renormalization group sense).  Consequently, as far as the low energy
physics is concerned, the spin and charge dynamics are completely decoupled
from each other.  Moreover, Luther and Emery showed that any space-time
electronic correlation function can be expressed as the product of a correlator
involving only the spin-fields and one involving only the charge fields. 

This work is profound, and although its importance was not widely appreciated
at the time, its influence has continued to grow ever since.  Under many
circumstances, the field theory methods are sufficiently powerful that all
relevant correlation functions, even for finite temperature, frequency,
or momentum, can be computed explicitly and exactly.  In the last couple of
years it has even become possible to compute various spectral functions
exactly in cases in which the spectrum of either the charge or spin
excitations is gapped, so that the bosonized field theory is importantly
non-linear.\cite{massive}  

From a more modern perspective, this body of work
concerns itself with the physics of a system at, or in the vicinity of a
quantum critical point.  An amazing thing about the 1DEG is that, due to the
special character of quantum fluctuations in one dimension, there are quantum
critical {\em phases}, not just critical points.  The successfulness of this
approach  is related to the fact that critical theories are conformally
invariant, and conformal invariance in one dimensional quantum or two
dimensional classical statistical mechanics turns out to strongly constrain the
nature of the correlation functions\cite{conformal}.

While the thrust of the present discussion concerns the broader implications
of electron fractionalization, it is worth noting that in recent years, literal
realizations of the 1DEG have become available for experimental study  in
nanowires\cite{nano}, buckeytubes\cite{bucky}, and edge states\cite{edge} in
quantum Hall systems.

\section{The SSH Model of Solitons in Polyacetylene}

The story of the soliton theory of polyacetylene in some sense starts with 
a beautiful result obtained in the context of relativistic quantum  field
theories by Jackiw and Rebbi\cite{jackiw}.  They noted that in certain
field theories, especially in one spatial dimension, one can find solitons
with fractional fermion number.  However, the subject really got started some 
years later with the discovery of the conducting polymer, 
trans-polyacetylene, by Shirakawa, MacDiarmid, and Heeger.  (There is no need
to review the importance of this discovery here.  However, readers who are
unfamiliar with polyacetylene are referred to the Fig. 1 and its caption,
for an introduction to this polymer.)  In thinking about the remarkable
properties of polyacetylene, Su, Schrieffer, and Heeger\cite{ssh} (SSH) (and
others\cite{others}) discovered that  solitons, which are precisely analogous
to those of  Jackiw and Rebbi, occur as the lowest energy electronic
excitations in the simplest imaginable model (the SSH model) of a set of
noninteracting electrons coupled to a lattice deformation (acoustic phonon). 
Here, however, the solitons have reversed charge-spin relations - they can be
neutral with spin 1/2, or if  spinless they have charge $\pm e$.   

Specifically, what SSH showed is that when an electron is added to an
otherwise neutral polyacetylene chain, it can break up into two pieces, one of
which carries the electron's charge and the other its spin.  This
observation bears a clear family relation with the phenomenon of spin-charge
separation in the 1DEG discussed above, but it is not identical.  In the first
place, undoped polyacetylene is a semiconductor, with a moderately large
($2\Delta\sim 2$eV) gap, and so is nowhere near being quantum critical.  The
solitons in polyacetylene are not low energy excitations;  the soliton creation
energy is approximately\cite{tlm} $E_s=2\Delta/\pi$.  Indeed, spin-charge
separation does not occur in the SSH model in the same precise sense as it
does in the 1DEG - for instance, there are substantial attractive interactions
between the charged and neutral solitons, which lead to a bound-state (a
polaron\cite{su}), with a binding energy $E_p - 2E_s =
- (2/\pi)[2 - \sqrt{2}]\Delta$.  In other words, spin charge coupling implies
that the lowest energy excitation of the system with the quantum numbers of an
electron is a quasi-particle with the same quantum numbers as the electron. 
In this sense, fractionalization is, in fact, a high energy feature of the
spectrum of the SSH model.  The lowest energy excitation made by
adding two electrons to the system is  a pair of charged solitons.  (
One can understand this effect qualitatively from the Hamiltonian
in Eq. (1);  under circumstances in which there is a gap to both spin and
charge excitations, the renormalization group flows carry the system to a
strong coupling fixed point where the terms in $H_{irr}$ are no longer
irrelevant, and in particular can lead to a  short-range attraction between
spin and charge solitons.)

Polyacetylene is a very complicated material, from a physicist's viewpoint.  It
is rather disordered, even when undoped, and doping introduces all sorts of
additional levels of randomness.  It is moderately one-dimensional, with an
in-chain bandwidth of order $W\approx$10eV as compared to an interchain
bandwidth of order one or two tenths of an eV;  this is not, however,
sufficient to permit truly one dimensional physics to be manifest at very long
distances, and clearly leads to rather strong soliton confinement. 
Nevertheless, many rather spectacular predictions of the soliton theory, both
spectroscopic and dynamical, were confirmed by experiments  performed with
great vigor, determination, and creativity by a large community of
scientists.  I think by now there can be no question that the basic features
of the soliton model of polyacetylene are not only right, in theory, but
applicable to a range of experiments, as well\cite{hkss}.

However, the real importance of the soliton model of polyacetylene was that it
introduced a new paradigm into the field.  It has led to ideas which have
had a lasting impact on condensed matter physics:
\begin{itemize}
\item{a)}  That there can exist quasi-particles
with fractional quantum numbers ({\it i.e.} quantum numbers that are unrelated
to those of an integer number of electrons and holes) that are robust
entities,  not just in low energy assymptopia, but in the
real-world realm of materials physics;
\item{b)}  That such quasi-particles have a topological character, from which
their stability derives;
\item{c)}  That these fractional quantum numbers
are sharp quantum observables\cite{sharp}, every bit as real as the charge of
the electron.
\end{itemize}

\subsection{Aside: the microscopic theory of polyacetylene}
Before I leave the subject of polyacetylene, there is one further point, not
directly related to fractionalization, that I will touch upon.  While the
character of the solitons, and the separation of charge and spin are robust
consequences of the broken symmetry (dimerized) groundstate, there are many
aspects of the SSH solution that are more microscopic, and model dependent. 
In part, it was the success of some of these more delicate aspects of the
theory that led to the general acceptance of the soliton model.  

In particular, there was a
remarkable quantitative {\em prediction} made by SSH, concerning the magnitude
of the dimerization in trans polyacetylene\cite{ssh}.  At the time of their
first work on the subject, the in-chain bandwidth,
$W$, was known within $20\%$ from various quantum chemical and band-structure
calculations.  Similarly, the spring constant ($K\approx
20$eV/$\AA^2$) of the
$\sigma$ bonds was well known to depend only on local chemistry, and so was
known reliably.  Thus, the only free parameter in the model was the
electron-phonon coupling constant, $\alpha$.  SSH determined the value of this
parameter, empirically, by fitting to the observed optical absorption gap, and
were then able to predict the magnitude of the lattice dimerization,
$u=0.04\AA$, on the basis of this fit.  Later quantitative measurements\cite{u}
found $u=0.03\pm 0.01\AA$, thus confirming this prediction.  

There are several things that are remarkable about
this triumph of the SSH theory. The first is that this is a one dimensional
system, so one might think that mean-field theory, which they employed, is
unreliable.  The neglect of electron-electron interactions seems, at first
sight, to be equally suspect, both because these interactions are known, from
quantum chemical studies of small polyenes, to be strong, and, as discussed
above, interactions in the 1DEG are known to completely destroy
Fermi-liquid behavior.  However, this success was not an accident.  
The fact that the phonon frequencies, $\hbar\omega_0$, are small on electronic
scales,
$\hbar\omega_0/\Delta \sim 10^{-1}$, can easily be shown to justify the
mean-field theory - indeed, effects of quantum lattice fluctuations were
systematically studied and found to be quite mild\cite{hirsh}.  The fact that
a weak-coupling theory is reasonable is guaranteed by the fact that the gap is
small compared to the bandwidth,
$\Delta/W \sim 10^{-1}$.  

The proof that the effects of electron-electron interactions are largely
perturbative was contained in some work I did with G.Zimanyi and
A. Luther\cite{zkl} (ZKL), long after it ceased to be a hot issue.
What  ZKL (and, at about the same time, J.Voit, as well\cite{voigt}) did was
to treat the problem of the 1DEG with both electron-electron (instantaneous)
repulsions and electron-phonon induced (retarded) attractions using standard
weak-coupling (one loop) renormalization group methods.  What we found was
that, even if at the microscopic level the electron-electron interactions are
much stronger than the electron-phonon interactions, so long as
$W/\hbar\omega_0$ and $W/\Delta$ are sufficiently large, the effective low
energy theory is always dominated by the electron phonon interactions.  The
electron-electron interactions have the effect of renormalizing the effective
electron-phonon coupling (curiously, they lead to a strong {\em enhancement}
of $\alpha$), but other than that, have only perturbative effects on the low
energy physics. Moreover, this effect is most pronounced when the electron
density is commensurate, or nearly so, {\it i.e.} when there is roughly
one electron per site.

Not only does this result justify, a
posteriori, the remarkable physical insight of SSH, it also explains why the
empirically determined value of $\alpha$ is large compared
to those found in microscopic, quantum chemical calculations;  the empirical
$\alpha$ is a renormalized coupling.  It also explains the
remarkable fact that when polyacetylene is ``overdoped,'' {\it i.e.} when the
electron concentration deviates by more than about 6\% from one
$\pi$-electron per carbon, it behaves like a nearly non-interacting
metal\cite{heegerandme};  due to the strong doping dependence of the effective
electron-phonon coupling, the expected Peierls instability of this
quasi-one-dimensional metal is suppressed\cite{salkola} to immeasurably small
temperatures.

\section{Fractionally charged Laughlin Quasi-particles}

From the first, the fact of electron fractionalization in the 1DEG and the
SSH model of polyacetylene was recognized as incompatible with
the conventional (Fermi liquid theory based) paradigms of condensed matter
physics.  However, it was widely believed at the time that this  was still
nothing more than a one-dimensional curiosity;   there is a general
theorem\cite{theorem} to the effect that solitons in a local field theory in
more than one dimension have infinite creation energy, unless coupled to a
suitable gauge field.  Thus, it was generally expected that:  1) 
No ``real'' ({\it i.e. } two or three dimensional) electronic system would ever
exhibit true electron fractionalization.  2)  The solitons of the
one-dimensional theory would be confined into integer charged multiplets the
moment interchain interactions were introduced.  Both of these statements turn
out to be wrong.

Almost immediately following the spectacular discovery of the fractional
quantum Hall effect\cite{tsui}, Laughlin\cite{laughlin} wrote down his famous
wavefunction, which captures the essential physics of this new
state of matter. An essential feature of this physics is that the
quasi-particle excitations of the fractional quantum Hall liquid are
vortex-like excitations with fractional charge.  The fact that the fractional
quantum Hall effect necessarily implies the existence of fractionally charged
quasi-particles can be seen straightforwardly from a slightly modified
version\cite{sondhireview} of the remarkable thought experiment introduced by
Laughlin in his original paper:

A fractional quantum Hall liquid is an incompressible state ({\it i.e.} it has
a gap, $\Delta$) with a quantized Hall conductance
\be
\sigma_{H}=(e^2/h) \ \ \nu^*
\ee
where $\nu^*$ is one of a set of discrete, rational fractions, of which the
most prominent is $\nu^*=1/3$.  Consider taking a system in its ground
state which exhibits the fractional quantum Hall effect, and piercing it with
an infinitesimal hole through which a magnetic flux, $\Phi$, can be threaded. 
Because the system has a gap, this can be done adiabatically, so long as the
rate at which the flux is threaded is slow compared to $\hbar/\Delta$.  Once
$\Phi$ reaches a magnetic flux quantum, $\Phi_0=hc/e$, the flux can be
effectively removed from the Hamiltonian by a gauge transformation;  the
resulting state is an eigenstate of the original Hamiltonian.  This
new eigenstate, however, has a well defined charge which has accumulated in the
vicinity of the hole.  To compute the induced charge, $Q$, consider measuring
the time integrated flux of current through a large ring which encircles the
hole
\be
Q=\int dt \dot Q = \int dt \oint d\vec l \times \vec J.
\ee
Now, the current density, $\vec J$, is the response,
$J_j =  \epsilon_{ij} \sigma_H E_j$  of the quantum Hall
state to the  electric field $\vec E(\vec r)= c^{-1}\dot{ \vec A}= \hat r \dot
\Phi/(2\pi r)$ produced by the time varying flux.  
(Here $\epsilon_{ij}$ is the Levi-Civita symbol.)  Consequently,
\be
Q=\sigma_H \Phi_0 = e\nu^*.
\label{Q}
\ee
If $\nu^*$ is fractional, there must exist quasi-particles with
fractional charge $e^*$ such that $ne^*=Q$, where $n$ is an integer!

The fractionally charged quasi-particles are still topological, in a sense, but
in a sense which escapes the general theorem.  Their topological character is
manifest as fractional statistics of the Laughlin quasi-particles under
exchange.  This was first recognized by Halperin\cite{halperin}, from a study
of the analytic properties of multi-quasi-particle wavefunctions, and later
derived in a more physical manner by Arovas, Wilczek, and
Schrieffer\cite{aws}, from a calculation of the Berry's phase when two
quasi-particles are adiabatically exchanged.  

As the above argument suggests, both the fractional
charge and the fractional statistics of the quasi-particles are, in a sense,
implicit in the fractional quantum Hall effect
itself\cite{manyanyon,laughlinnobel,laughscience}.  The 
fractional quantization of the Hall conductance has been observed with an
accuracy of better than one part in 10$^{-4}$.  Experiments designed to
directly measure\cite{vladimir} the fractional charge of the
quasi-particle have been carried out, with results consistent with
expectations, although with nowhere the same level of accuracy.  Experiments
have been proposed\cite{chamon} to directly measure the fractional statistics,
as well, although they will be hard.  However, there is no question in anyone's
mind\cite{laughlinnobel} that the Laughlin quasi-particles exist, that they
are as robust as the fractional quantum Hall state itself, and that they have
the predicted fractional charge and statistics.

One subtlety, with potentially significant consequences, is swept under the
rug in the above discussion.  It is possible to find circumstances in which,
depending on details of the electronic Hamiltonian, there can be
more than one possible quantum Hall liquid with the same value of the Hall
conductance.  However, these different states generally will have
quasi-particles with different quantum numbers, although always consistent
with the constraint that there exist multiplets with charge $Q$ given in Eq.
\ref{Q}.  The macroscopic distinctions between such states can be classified
in terms of  the quantum numbers of the quasi-particles.  Alternatively, they
can be classified in terms of certain topological properties of the ground
state\cite{wentopo}, namely the ground-state degeneracies on closed surfaces
of varying connectivities.  There is, of course, an intimate connection between
these two approaches.  It would seem that the former is more closely related
to a conceivable experiment, although the topological structure of the ground
state is also related to the character of the edge states
produced in a system with boundaries.  

\section{RVB and spin-charge separation in two dimensions}

The quantum Hall system is still very special.  It is two dimensional\cite{3d},
but so is the electron gas in real MOSFETS and heterojunction devices, among
other systems.  However, the large magnetic field explicitly breaks
time-reversal and reflection symmetry.  Thus, although following Laughlin's
work, there was no denying that fractionally charged particles were a feature
of the ``real'' world, they still occupied a small corner of that world.

Immediately following the discovery of high temperature
superconductivity\cite{mueller}, Anderson proposed\cite{pwa87} that the key
to the problem lay in the occurrence of a never before documented state of
matter, a spin-liquid or ``resonating valence bond'' (RVB) state, related to a
state he originally proposed\cite{fa} for quantum antiferromagnets on a
triangular (or similarly frustrating) lattice.  A spin-liquid, in this
context\cite{rvbreview}, is defined to be an insulating state (with a charge
gap) and an odd integer number of electrons per unit cell which breaks neither
spin-rotational nor translational symmetry.  Following Anderson, Jim Sethna,
Dan Rokhsar and I showed\cite{krs} that a consequence of the existence
of such a spin-liquid state is that there exist quasi-particles with reversed
charge spin relations, just like the solitons in polyacetylene:  charge 0 spin
1/2  ``spinons'' and charge e spin 0 ``holons.''  Indeed, these quasi-particles
were recognized as having a topological character\cite{krs,laughscience}
analogous to that of the Laughlin quasi-particles in the quantum Hall effect. 

There was a debate at the time concerning the proper exchange statistics, with
proposals presented identifying the holon as a boson\cite{krs,bza}, a
fermion\cite{read}, or a semion\cite{kalmeyer}.  I now
believe that all sides of this debate were correct, in the sense that there is
no universal answer to the question - depending on details of the
Hamiltonian, it is possible to imagine\cite{holonstatistics,kimnayak}
transitions occurring between states in which the holon has different
statistics.  In fact, this debate represented the first steps in the
theoretical exploration of the topological structure of spin liquid
states.  We will return to this in Sec.
\ref{topological}, below.  

The real question-mark hanging over the whole subject, as was pointed out most
forcefully by Read and Sachdev\cite{rs}, is whether, and under what
circumstances a spin-liquid exists;  indeed, they presented strong arguments
that the most straightforward quantum disordering of an antiferromagnet will
lead to a spin-Peierls state, rather than a spin-liquid.  Moreover, for more
than a decade, despite extensive effort, no one succeeded in producing a model
system which could be convincingly shown to exhibit a spin-liquid
ground-state\cite{rvbreview}.  

In this context, I am happy to report that very recently, Moessner  and
Sondhi\cite{shivaji} have managed to do just this!  They have considered
a model\cite{rk} on a triangular lattice (thus returning very closely to the
original proposal of Anderson) which is a bit of a
caricature in the sense that the constituents are not single electrons, but
rather valence bonds (hard-core dimers), much in the spirit pioneered by
Pauling. However, the model is sufficiently well motivated, microscopically,
and the spin-liquid character sufficiently robust, that I believe it is
reasonable to declare victory.  Moreover, the resulting spin-liquid state
does not break time-reversal symmetry.  This means that the point of principle
has been established.  Spin charge separation (and,
presumably, still more exotic forms of electron fractionalization) {\em can}
occur in more than one dimension, and in the absence of either explicit or
spontaneous time-reversal symmetry breaking!  It is now only a matter of time
until this phenomenon is either confirmed in some existing material (maybe even
the high temperature superconductors), or discovered in a new system.

\section{Fractionalization at quantum critical points}

Electron fractionalization of a sort can occur at quantum critical points in
systems between their upper and lower critical dimensions, in the sense that
an injected electron will, with probability one, decay into a multi-particle
continuum\cite{sachdev}.  However, this type of electron
fractionalization is quite different from the electron fractionalization
we have been discussing.  In the first place, in more than one dimension,
and in the limit $T\rightarrow 0$,
quantum critical phenomena typically occur at a critical {\em point}, as
opposed to in a critical {\em phase} of matter.  In the second place,
there is no known quasi-particle description of the elementary excitations of
such a quantum critical system\cite{wilson,subirbook,chn};  in particular,
there is no currently accepted\cite{laughlinbold} description of the critical
state in terms of excitations with fractional quantum numbers.

However, there is another recent development which is exotic, but very
intimately connected to the type of electron fractionalization discussed
here.   It was recognized\cite{lubensky} a couple of years ago that a certain
class of layered (i.e. quasi-two dimensional) {\it classical} systems can
exhibit a critical {\em phase}, a phase, moreover,  in which at low energy, the
couplings between the layers are negligible.  Such ``floating'' phases may, in
fact, have been seen in lamellar DNA-lipid bilayer complexes\cite{DNA}. 
Moreover, a similar phenomenon can occur in the quantum theory of quasi-one
dimensional systems\cite{nature,prl,carpentier,kane,kun}.  In such systems,
even though at a microscopic level there are weak but non-vanishing couplings
between chains, all interchain couplings are irrelevant in the renormalization
group sense; the low energy physics of the system is that of a set of
decoupled 1DEGs!

Among other things, this means that precisely the same form of one
dimensional spin-charge separation discussed in the first section occurs in
such systems, even though at a microscopic level there are finite higher
dimensional couplings.  For the models studied to date, the floating
phases occur only in an extreme region of parameter space\cite{kanelub}.  It
may be that this implies that floating phases are very rare, indeed.  However,
now that we know they exist in principle, we can begin looking for them. 
Already, there is evidence\cite{uchida} that such a floating phase occurs in
one particular member of the family of high temperature superconductors.    

\section{Where else might electron fractionalization occur?}

I think this is the tip of the iceberg.  The Fermi liquid based view of the
electronic properties of solids has been very successful as a basis for
understanding the essential physics of a wide range of conventional solids,
including metals, semiconductors, and superconductors.  It runs into
difficulty in highly correlated solids, where so called ``non Fermi
liquid'' (NFL) behavior is observed;   here, even the given name admits to
our complete lack of any successful theory of what causes this behavior.  At
best, the Landau quasi-particles are very strongly interacting in such
solids.  More likely, in many cases no such quasi-particle description is
possible, and in some cases, electron fractionalization provides the correct
framework for thinking about these systems.

\subsection{Electron fractionalization as intermediate scale physics}

While the conditions for true electron fractionalization, in the sense that it
is really possible to have arbitrarily widely separated excitations with
fractional quantum numbers, may be rather restrictive,
the notion may be much more widely applicable.  

A Fermi liquid (as
contrasted with a Boltzman liquid) is only precisely defined in the limit
that the temperature,
$T\rightarrow 0$;  however, systems with a true Fermi liquid ground-state are
certainly rare, at best, and may well (due to the famous Kohn-Luttinger
theorem\cite{shankar}) be non-existent.  Never-the-less, over a broad
intermediate range of temperatures such that $T_F \gg T  \gg T_c$, the Fermi
liquid description is both qualitatively and quantitatively valid. (Here
$T_F$ is the Fermi temperature and
$T_c$ is some ordering temperature, for instance to a superconducting state.)  

Similarly, there occur circumstances in which a
fractionalized description of the spectrum of a system over an intermediate
range of temperatures, frequencies, and length scales may be valid, even though
in assymptopia it breaks down.  This is certainly true of a typical quasi-one
dimensional system where, at temperatures above any ordering temperature, the
system can be treated as one dimensional, with the effects of interchain
coupling making only perturbative corrections to the 1D physics.  Even below
$T_c$, the physics of the 1DEG is manifest at all but the lowest frequencies. 
Recently, considerable progress\cite{erica,dror,cortney,mpanayak} has been
made in obtaining an understanding of the consequences of electron
fractionalization for various electronic spectroscopies of correlated
electronic systems, and on the nature\cite{erica,cortney,zaanen} of the
crossover from a fractionalized spectrum above $T_c$ to a spectrum with a well
defined quasi-particle piece below
$T_c$.  In many respects, these spectra account\cite{erica,dror} well for
properties of the observed spectra (especially those obtained in angle
resolved photoemission\cite{arpes}) in the high temperature superconductors,
including many features that simply cannot be understood in the context of
Fermi liquid theory. It has also been suggested\cite{sudipandme}, that
spin-charge separation at intermediate scales may generally be the basis of
an electronic mechanism of high temperature superconductivity.

\subsection{Topological order and electron fractionalization}
\label{topological}

We now address the problem of classifying phases in
which true electron fractionalization occurs, {\it e.g.} in which
spinons are deconfined.  It is now clear from the work of  Wen\cite{wentopo}
and Senthil and Fisher\cite{senthil} that the best macroscopic characterization
of fractionalized phases in two or more dimensions, given that they frequently
posses no local order parameter, is  topological.  Specifically, a
fractionized phase  exhibits certain predictable ground-state degeneracies on
various closed surfaces - degeneracies which Senthil and Fisher have given a
physical interpretation in terms of ``vison expulsion.''  Unlike the
degeneracies associated with conventional broken symmetries, these
degeneracies are not lifted by small external fields which break either
translational or spin rotational symmetry.  It has even been
shown\cite{holonstatistics,sachdevsenthil,senthil} (funny as this may sound)
that  topological order is amenable to experimental detection.    Once 
topological classification is accepted, the one-to-one relation between
spin-liquids and electron fractionalization, implied in our previous
discussion,  is eliminated.  Indeed, it is possible to
imagine\cite{senthil,mpanayak} ordered (broken symmetry) states, proximate to
a spin liquid phase, which will preserve the ground-state degeneracies of the
nearby spin liquid, and hence will exhibit spin-charge separation.

\subsection{Electron fractionalization and quantum computing}

One of the most exciting areas where electron fractionalization may play an
important role is in the developing field of quantum computing.  It was
recently shown\cite{freidman} that many of the vexing problems of decoherence,
which are barriers to construction of a functioning quantum computer, can be
avoided if the quantum states in question have an appropriate topological
character.  Certain types of electron fractionalization, including those which
occur in some complicated quantum Hall states\cite{paired} and in the RVB
spin-liquid\cite{kirill}, have topological structure that could be useful for
this purpose.  Here, even if (as seems likely) rather special circumstances are
required to obtain true electron fractionalization, it may be worthwhile
to seek artificial methods for achieving those special circumstances.

\begin{figure}[bht]
\narrowtext
\begin{center}
\leavevmode
\noindent
\hspace{0.3 in}
\centerline{\epsfxsize=2.6in \epsffile{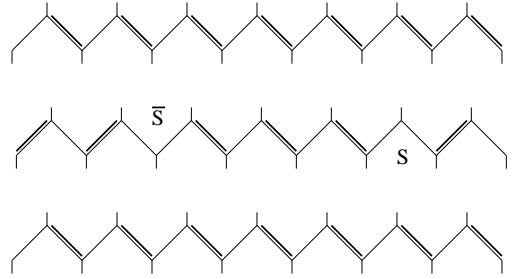}}
\end{center}
\caption 
{Schematic Representation of Solitons in Polyacetylene:  The figure represents
three polyacetylene chains with a soliton (S) anti-soliton ($\bar S$) pair on
the middle one. In terms of structure, the vertices indicate the position of C
atoms and the ends of the vertical bonds indicate H positions;  the double
bonds are slightly shorter than the single bonds.  Note that there are two
degenerate patterns of alternating single and double bonds (dimerization) in
the ground-state;  we talk of a given pattern breaking the translational
symmetry of the polymer, but in fact, because of the zig-zag, it is actually a
reflection plane symmetry (about a C site) which is being broken, or
alternatively a screw symmetry.   A strong coupling characature of the
electronic state can be deduced from the figure as follows:  Associate with
each C atom the two core 1S electrons.  The thin lines represent a pair of
electrons in a bonding $\sigma$ orbital, and the heavy lines a pair of $\pi$
electrons (that is, electrons in an out-of-plane 2P C orbital) in a bonding,
or valence bond, state.  The local neutrality of the pristine (outer)
polyacetylene chains is easily seen from the fact that there are 2 core
electrons, 3 $\sigma$ electrons and 1 $\pi$ electron per C, and 1 $\sigma$
electron per H.  From this it is clear that S and $\bar S$ each have charge
$+e$ and, since all electrons are paired, spin 0.  There is also, clearly, a
non-bonding $\pi$ orbital left over on the central C in each soliton.  If this
orbital were singly occupied, the soliton would have charge 0 and spin 1/2,
while if it were doubly occupied, it would have charge $-e$ and spin 0. 
Although in reality the
$\pi$ electrons, in particular, are rather extended, so that the $\pi$
electron density on the short bonds is only a few percent larger than that on
the long bonds, because the spectrum is gapped, the quantum numbers of the
solitons are unchanged upon adiabatic continuity from the strong coulpling
limit, described above.  All the fancy topological theorems that have been
applied to this problem are no better than this simple derivation, often
called ``the Schrieffer counting argument.''  The fact that interchain
couplings inevitably cause soliton confinement can also be seen from the
figure.  Between the solitons, the dimerization on neighboring chains has the
same sense, while beyond them, it has the negative sense.  Since interachain
couplings favor one relative phase or the other by a given energy per unit
length of chain, there is manifestly a confining potential which grows
linearly with the separation between solitons, if the interchain couplings are
not zero.}
\label{fig:1}
\end{figure}

\noindent{\bf Acknowledgements:}  Substantial improvements in
this  manuscript were made in response to comments by M.P.A.Fisher, E.Fradkin,
C.Nayak and S.Sachdev.
   

\end{multicols}


\begin{references}
\bibitem{kanelub}  There are some subtleties and caveats associated with this
statement.  For a discussion of the current state of the theory of this issue,
see Ref. [\ref{kane}].
\end{references}
\end{document}